\documentclass[12pt,superscriptaddress,runinaddress,floatfix,preprintnumbers,showpacs,nofootinbib]{revtex4-1}
\usepackage{amssymb,amsmath}
\usepackage[dvips]{graphicx}
\usepackage{color}
\usepackage{dcolumn}
\usepackage{amssymb}
\usepackage{amsmath}
\usepackage{amsthm}
\usepackage{latexsym}
\usepackage{graphicx}
\usepackage{color}
\usepackage{ulem}
\usepackage{hyperref}
\usepackage{fullpage}
\usepackage{setspace}
\usepackage{multirow}
\usepackage{subfloat}
\usepackage{subfigure}
\usepackage{cancel}
\usepackage{listings}
\usepackage{pgf}
\usepackage{tikz}
\usetikzlibrary{shapes,arrows}
\usepackage{verbatim}%
\usepackage[margin=1.5cm]{geometry}

\newcommand{\be}{\begin{equation}}
\newcommand{\ee}{\end{equation}}
\newcommand{\ba}{\begin{array}}
\newcommand{\ea}{\end{array}}

\tikzstyle{decision} = [diamond, draw, fill=blue!20, 
    text width=4.5em, text badly centered, node distance=3cm, inner sep=0pt]
\tikzstyle{block} = [rectangle, draw, fill=blue!20, 
    text width=5em, text centered, rounded corners, minimum height=4em]
\tikzstyle{block2} = [rectangle, draw, fill=red!20, 
    text width=5em, text centered, rounded corners, minimum height=4em]
\tikzstyle{block3} = [rectangle, draw, fill=green!20, 
    text width=6em, text centered, rounded corners, minimum height=4em]
\tikzstyle{line} = [draw, -latex']
\tikzstyle{cloud} = [draw, ellipse,fill=red!20, node distance=3cm,
    minimum height=2em]
\lstdefinestyle{nonumbers}
{numbers=none}
\lstdefinestyle{custom}{
  language=c++,
  numbers=none,
  stepnumber=1,
  numbersep=10pt,
  tabsize=4,
  showspaces=false,
  showstringspaces=false
}

\begin{document}
 \begin{flushright}
    ADP--16--42/T998\\
    CoEPP--MN--16--29
  \end{flushright}
\title{Dark matter candidates in the \\ constrained Exceptional Supersymmetric Standard Model\\[4mm]}
 
\author{
P.~Athron${}^{a,1}$,
A.W.~Thomas$^{b}$,
S.J.~Underwood$^{b,2}$,
M.J.~White$^{b,3}$
\\[9mm]
{\small\it $^a$ ARC Centre of Excellence for Particle Physics at the Terascale,}\\
{\small\it School of Physics \& Astronomy, Monash University, Melbourne VIC 3800, Australia}\\[3mm]
{\small\it $^b$ ARC Centre of Excellence for Particle Physics at the Terascale and CSSM,}\\
{\small\it School of Physical Sciences, The University of Adelaide, Adelaide SA 5005, Australia}\\[3mm]
{\small\it $^1$ peter.athron@coepp.org.au}\\
{\small\it $^2$ sophie.underwood@adelaide.edu.au}\\
{\small\it $^3$ martin.white@adelaide.edu.au}
}

\begin{abstract}
\begin{center}
{\bf Abstract}
\end{center}
The Exceptional Supersymmetric Standard Model (E$_6$SSM) is a low
energy alternative to the MSSM with an extra $U(1)$ gauge symmetry and
three generations of matter filling complete 27-plet representations
of $E_6$.  This provides both new D and F term contributions that raise
the Higgs mass at tree level, and a compelling solution to the
$\mu$-problem of the MSSM by forbidding such a term with the extra
$U(1)$ symmetry.  Instead, an effective $\mu$-term is generated from the VEV of an
SM singlet which breaks the extra $U(1)$ symmetry at low
energies, giving rise to a massive $Z^\prime$. We explore the
phenomenology of the constrained version of this model (cE$_6$SSM) in
substantially more detail than has been carried out previously,
performing a ten dimensional scan that reveals a large volume of
viable parameter space.  We classify the different mechanisms for
generating the measured relic density of dark matter found in the
scan, including the identification of a new mechanism involving mixed bino/inert-Higgsino dark matter.  We show which mechanisms can evade the latest direct detection
limits from the LUX 2016 experiment.  Finally we present benchmarks
consistent with all the experimental constraints and which could be
discovered with the XENON1T experiment.
\end{abstract}
\maketitle

\newpage
\section{Introduction}

With the discovery of a $125$ GeV Higgs boson, all elementary particles in the
Standard Model (SM) of particle physics have been discovered and the
model is extremely well verified as a description of nature, fitting
observations from past and current collider experiments.  However, the
SM cannot explain the observed dark matter, which
constitutes 23\% of the universe's mass-energy content and has
motivated many proposed modifications to the SM. Supersymmetric extensions in particular, although motivated for
many other reasons, are also often favoured for providing viable
Weakly Interacting Massive Particle (WIMP) candidates for dark
matter. For instance, the application of R-parity, a $Z_2$ symmetry
meant to preserve baryon and lepton number, to the Minimal
Supersymmetric Standard Model (MSSM) ensures the stability of the
lightest supersymmetric particle.

However the MSSM now requires considerable fine tuning to obtain a Higgs mass
of $125$ GeV and it has a so-called ``$\mu$ problem'' associated with it. The coupling
between the Higgs superfields, $\mu$, is the only dimension one
parameter in the MSSM superpotential.  Since, for phenomenological
reasons, $\mu$ should be of the same order of magnitude as the
electroweak (EW) scale, despite there being no physical connection
between them, this presents a naturalness problem.

Here we investigate dark matter in a well motivated E$_6$-inspired
model. We explore the different types of neutralino dark matter that
can explain the observed relic density, while satisfying collider
constraints and examine the impact of recent direct detection
experiments on the model. $E_6$-inspired supersymmetric
models \cite{E6-review,E6} provide a solution to the $\mu$ problem
wherein the $\mu$-term is forbidden by an extra $U(1)$ gauge symmetry
which appears from the breakdown of $E_6$ and survives to low
energies, where it is broken close to scale of electroweak symmetry
breaking.  The break down of this extra $U(1)$ symmetry occurs when an
SM singlet picks up a VEV, dynamically generating an effective
$\mu$-term without the accompanying domain wall / tadpole problems
that appear in the NMSSM \cite{Ellwanger:2009dp,Maniatis:2009re}.

$E_6$-inspired models with an extra $U(1)$ gauge symmetry have
  attracted extensive interest in the literature
  \cite{Daikoku:2000ep,Kang:2004ix,Ma:1995xk,baryogen,g-2,Suematsu:1997tv,Suematsu:1997au,Keith:1996fv,GutierrezRodriguez:2006hb,Suematsu:1997qt,Ham:2008fx,E6neutralino-higgs,Asano:2008ju,Stech:2008wd,E6-higgs,Accomando:2010fz,Hesselbach:2001ri,Gherghetta:1996yr,Zprime,Kang:2007ib,Langacker:1998tc,Cvetic:1997ky,Suematsu:1994qm,Keith:1997zb,Athron:2015tsa}.
  Here we work specifically with a $U(1)_N$ gauge symmetry at low
  energies under which the right handed neutrino remains
  interactionless. This is used in the Exceptional Supersymmetric
  Standard Model (E$_6$SSM)
  \cite{King:2005jy,King:2005my,brief-review} and closely related
  models \cite{Howl:2007zi, Braam:2009fi, Braam:2010sy,
    Nevzorov:2012hs, Nevzorov:2013ixa, Athron:2014pua} to allow
  right-handed neutrinos to gain mass far above the TeV scale and
  trigger a see-saw mechanism that explains the tiny observed masses
  of neutrinos.  This can also provide a leptogenesis mechanism to
  explain the baryon asymmetry in the Universe
  \cite{Hambye:2000bn,King:2008qb}.


The gauge coupling running in the E$_6$SSM at the two-loop level leads
to unification more precisely than in the MSSM \cite{King:2007uj},
while in slightly modified scenarios two-step unification can take
place \cite{Howl:2007hq,Howl:2007zi}. If the exotic particles are
light in these models this can open up non-standard decays of the
SM--like Higgs boson
\cite{Hall:2010ix,Nevzorov:2013tta,Athron:2014pua}. In the constrained
version of the E$_6$SSM, the particle spectrum, collider signatures
and fine tuning have been studied
\cite{Athron:2008np,cE6SSM1,cE6SSM2,cE6SSM3,cE6SSM4,Athron:2013ipa}. The
threshold corrections to the $\overline{DR}$ gauge and Yukawa
couplings in the E$_6$SSM were calculated and the numerical impact in
the constrained version examined
\cite{Athron:2012pw,Voigt:2010ksa}. The impact of gauge kinetic mixing
in the case where both the extra U(1)'s appearing from the breakdown
of E$_6$ are present at low energy was studied in
\cite{Rizzo:2012rf}. The E$_6$SSM was also included in studies looking
at how first or second generation sfermion masses can be used to
constrain the GUT scale parameters \cite{Miller:2012vn} and the
renormalization of VEVs \cite{Sperling:2013-1, Sperling:2013-2}.  Very
recently the model has been studied in the context of electroweak
baryogenesis \cite{Chao:2014hya} and the possibility of it explaining
the recent apparent diphoton excess was also discussed
\cite{Chao:2016mtn,King:2016wep,Staub:2016dxq}.

The situation for dark matter in these models can be quite different
from that of the MSSM. The E$_6$SSM neutralino sector is extended
compared to the MSSM by the both extra matter fields and the fermion
component from the extra vector superfield associated the extra
$U(1)$.  If only this new gaugino and the third generation singlino
(superpartner of the singlet Higgs field which breaks the extra $U(1)$
symmetry) mix with the MSSM-like gauginos and Higgsinos, then the
neutralino sector would be that of the $U(1)$-extended Supersymmetric
Standard Model (USSM) \cite{Kalinowski:2008iq}.

However if one considers interactions from $27_i \times
27_j \times 27_k$ then the superpotential will have a term amongst the
Higgs-like fields analogous to that of the NMSSM, but with indices
running over all three generations,
\begin{equation}
\Sigma_{ijk} \lambda_{ijk} S_i H^d_{j} H^u_{k} \in W_{E6SSM}, \hspace{1cm} i, j, k \in (1,2,3).
\label{Eq:lambdaijk}
\end{equation}
As will be discussed later, the first two generations of Higgs-like fields will remain inert and will not develop a VEV, while the 3rd generation will be the actual Higgs fields, with the neutral scalar components developing VEVs. The effective $\mu$ parameter is then provided by $\mu_{eff} = \frac{s\lambda_{333}}{\sqrt{2}}$, where $s$ is the VEV for the singlet scalar field $S_3$.  

Such an interaction allows mixing between the Higgsinos and singlino
(i.e. the superpartners of the actual Higgs fields) and the ``inert''
Higgsinos and ``inert'' singlino which are the fermion components of
the inert first and second generation Higgs-like superfields. Indeed,
scenarios where the correct relic density can be obtained entirely
from the inert sector have been explored \cite{Hall:2009aj}. However
because the ``inert'' singlinos are always rather light states, these
scenarios are now ruled out by limits on non-standard Higgs decays and
direct detection of dark matter experiments such as LUX
\cite{Akerib:2013tjd,Akerib:2015rjg,Akerib:2016vxi}.  Dark matter has
also been studied in a related $E_6$ model where there is a single,
exact custodial symmetry that decouples all of the ``inert''
neutralinos from the USSM-like neutralino states, rendering the dark
matter situation much more similar to that of the MSSM
\cite{Athron:2015vxg, Athron:2016gor}.

In this article we instead consider specifically the EZSSM
\cite{Hall:2011zq} scenario, where only the light singlino states have
been decoupled from the rest of the neutralino sector and contribute
negligibly to the dark matter relic density. Only specific scenarios
with bino-like dark matter candidates have been examined for this
previously. In those scenarios the relic density is explained through a
new mechanism that involves the bino scattering off SM states into
inert-Higgsinos, where the latter need to have masses very close to that of
the bino for this to work.  However, we will show that the model has a
much richer set of possibilities for obtaining the measured relic density of
dark matter.

Here we expand substantially on previous work exploring the parameter
space of the E$_6$SSM \cite{cE6SSM1,cE6SSM2,cE6SSM3,cE6SSM4,Athron:2013ipa}.  For the first time we include the
relic density calculation in a systematic exploration of the parameter
space of the model.  Furthermore, we vary the full set of parameters in
the constrained model, rather than just a two or three dimensional
subset.  This includes varying $Z_2^H$ violating
Yukawa couplings that mix the exotic neutralino (i.e.~the
inert-Higgsino) couplings with the USSM sector neutralinos formed by
the bino, wino, Higgsinos and fermion components of the gauge and
singlet supermultiplets.

With this more systematic approach we reveal the different possible
neutralino dark matter scenarios that can explain the relic density.
We find that the dark matter candidate can be predominantly bino,
Higgsino or inert-Higgsino in nature, or it can have a significant
mixture of two or all three of these.  In particular, the scenarios
involving a significant inert-Higgsino dark matter admixture have not
been discussed before in any $E_6$-inspired model.  Scenarios with a
significant admixture of inert-Higgsino and bino are very interesting as
these scenarios can fit the relic density without driving the spin-independent cross-section up, as happens in the MSSM and $E_6$-inspired
models where the dark matter candidate has substantial admixtures of
Higgsino and bino.  
 
 We also show that it is possible to fit the relic density of dark
 matter simultaneously with collider data, such as the $125$ GeV
 Higgs mass and other limits from collider experiments, across a wide
 range of parameters.  We present new benchmarks from the scans which
 can do this and represent the different types of dark matter that we
 have identified.


This paper is structured as follows. In Section II we review the model
with particular focus on the neutralino sector. In Section III we
outline our scan procedure. In Section IV we show the results of scans
of the parameter space which reveal that a large volume of the
parameter space is consistent with all available data and the limits
from direct detection on the exotic couplings. We then identify the
characteristics of the new dark matter candidates and present a set of benchmark
points for scenarios that survive the latest limits from direct
detection experiments. Finally we present our conclusions in Section
V.
\section{E$_6$SSM}
The breakdown of $E_6$ can lead to two extra $U(1)$ gauge groups
defined by the breaking of $E_6\to SO(10)\times U(1)_{\psi}$, and the
subsequent breaking of $SO(10)$ into $SU(5)$, $SO(10)\to SU(5)\times
U(1)_{\chi}$ (this is reviewed in e.g.~\cite{Langacker:2008yv}). In
$E_6$-inspired models that solve the $\mu$ problem, one linear
combination survives to low energies and, in the E$_6$SSM, this combination is 
\be U(1)_N = \frac{1}{4}U(1)_{\chi} + \frac{\sqrt{15}}{4}U(1)_{\psi}.  \ee 

The full low energy gauge group of the E$_6$SSM is then \be
SU(3)_C\times SU(2)_W\times U(1)_Y\times U(1)_N. \ee This is subsequently broken down to $SU(3)_C\times U(1)_e$ when the Higgs fields that
couple to up-type fermions, $H_u$, down-type fermions, $H_d$, and the
singlet Higgs field, $S$, pick up VEVs.

The E$_6$SSM has an extended particle content to include three
complete $27_i$ representations of $E_6$ (where $i$ runs from 1 to 3). This ensures the cancellation of gauge anomalies in each generation.
The three families decompose as:
\begin{equation}
27_i \rightarrow (10,1)_i + (5^*,2)_i + (5^*,-3)_i + (5,-2)_i + (1,5)_i + (1,0)_i,
\end{equation}
where the first quantity in each bracket is the $SU(5)$ representation
and the second quantity is the extra $U(1)_N$ charge (the
decomposition occurs under a $SU(5) \times U(1)_N$ subgroup of
$E_6$). The first two terms contain quarks and leptons, the third and
fourth terms contain up- and down-type Higgs-like doublets $H^u_{i}$
and $H^d_{i}$ as well as additional exotic coloured states $D_i$ and
$\bar{D}_i$, the fifth contains the SM-singlet fields $S_i$ and the
last contains the right-handed neutrinos.

The matter content is then completed with the inclusion of two
additional $SU(2)$ multiplets $H^\prime$ and $\overline{H}^\prime$,
which are the only components from additional $27^\prime$ and
$\overline{27}^\prime$ that survive to low energies. These incomplete
multiplets at low energies ensure that gauge coupling unification can
be achieved. The low energy matter content of the model looks like,
\be \ba{c} (Q_i,\,u^c_i,\,d^c_i,\,L_i,\,e^c_i)
+(D_i,\,\bar{D}_i)+(S_{i})+(H^u_i)+(H^d_i) +H^\prime
+\overline{H}^\prime, \ea
\label{hd7}
\ee where $i=1,2,3$ runs over the three generations of $27_i$ and
corresponds to the traditional three generations of matter of the SM
and MSSM.

The actual Higgs fields that develop VEVs are $H_u := H^u_3$, $H_d :=
H^d_3$ and $S := S_3$. The remaining Higgs-like fields
$H^d_{\alpha}$, $H^u_{\alpha}$ and $S_{\alpha}$ (where $\alpha=1,2$
runs over the first two generations) do not develop VEVs and so are
referred to as ``inert'' Higgs bosons.

\subsection{The superpotential, $Z_2$ symmetries and soft masses}
The full superpotential that can arise from the $27_i \times 27_j \times 27_k$ interactions may be written as 
\begin{equation}
W_{E6} = W_0 + W_1 + W_2,
\end{equation}
where
\begin{eqnarray}
W_0 &=& \lambda_{ijk} S_i H^d_{j} H^u_{k} + \kappa_{ijk} S_i D_j \bar{D}_k + h^N_{ijk} N^c_i H^u_{j} L_k \nonumber \\
& & + h^U_{ijk} u^c_i H^u_{j} Q_k + h^D_{ijk} d^c_i H^d_{j} Q_k + h^E_{ijk} e^c_i H^d_{j} L_k, \\
  W_1 &=& g^Q_{ijk} D_i Q_j Q_k + g^q_{ijk} \bar{D}_i d^c_j u^c_k,\\
W_2 &=& g^N_{ijk} N^c_i D_j d^c_k + g^E_{ijk} e^c_i D_j u^c_k + g^D_{ijk} Q_i L_j \bar{D}_k.
\end{eqnarray} 
However, there are phenomenological problems with such a
superpotential, since at this point lepton and baryon number violating
operators that lead to rapid proton decay (an obviously undesirable
feature of any model) are not forbidden and there are also terms which
can lead to large flavour-changing neutral currents.

In the original formulation of the E$_6$SSM
\cite{King:2005jy,King:2005my}, the solution employed is to
impose two discrete symmetries. The first one is an analogue of
R-parity, which is either a $Z_2^L$ symmetry, where the superfields
which are odd under this symmetry are the set, $L_i, e_i^c, N_i^c,
H^\prime, \overline{H}^\prime$, or a $Z_2^B$ symmetry,
where the set of even superfields are extended to include the exotic
colored superfields $D_i$ and $\overline{D}_i$. If one assumes the
$Z_2^L$ symmetry then the interactions in $W_1$ are allowed and this
implies that the exotic coloured superfields are diquark in nature. If
one instead assumes $Z_2^B$ then they must be leptoquark in nature,
since the interactions in $W_2$ are allowed.

The second discrete symmetry is $Z^H_2$, under which $S_3$, $H^d_{3}$
and $H^u_{3}$ are even while every other field is odd. As a consequence, any term in
the superpotential that violates $Z^H_2$ (by containing superfields
adding up to a net odd value) is forbidden. However, the $Z^H_2$
symmetry cannot be exact, since it forbids all terms that would
otherwise allow for the decay of exotic quarks. Therefore in the
standard approach there is an approximate $Z^H_2$ symmetry. Although
this may seem rather {\it ad hoc}, it is worth noting that family symmetries
can lead to symmetries which operate in effectively the same way as the
approximate $Z_2^H$ symmetry introduced for phenomenological reasons
\cite{Howl:2009ds}. Alternatively, an exact custodial symmetry may be
used \cite{Nevzorov:2012hs}.

 The couplings $\lambda_{ijk} S_i H^d_{j} H^u_{k}$,
 which were highlighted in the introduction (Eq.~\ref{Eq:lambdaijk}), are
 affected by this symmetry. The following
 couplings are suppressed by this symmetry:
\be
x_{d\alpha} := \lambda_{33\alpha}, \;\;\; x_{u\alpha} := \lambda_{3\alpha3}, \;\;\; 
z_{\alpha} := \lambda_{\alpha 33}, \;\;\; c_{\alpha\beta\gamma} := \lambda_{\alpha\beta\gamma},   
\ee where $\alpha, \beta, \gamma \in {1,2}$ runs over the inert generations of the Higgs-like states. This leaves just \be \lambda := \lambda_{333}, \;\;\; \lambda_{\alpha\beta} := \lambda_{3\alpha\beta}, \;\;\;  f^d_{\alpha\beta} := \lambda_{\alpha3\beta} \;\;\; f^u_{\alpha\beta} := \lambda_{\alpha\beta3} \ee as unsuppressed couplings.

However, in order to ensure that only the third generation Higgs-like fields
acquire VEVs, large Yukawa couplings should not appear in
renormalization group equations (RGEs) for the soft masses of the
first two generations of Higgs-like fields.  As a result the $f^u$ and
$f^d$ couplings cannot be so large and this implies that the singlinos
are then always very light states since they get their masses from
these interactions when $H_u$ and $H_d$ get VEVs.

This means that the inert-singlinos are always the lightest neutralino
states. It is possible that these inert states can explain all of the
observed dark matter \cite{Hall:2009aj}. However, constraints from
direct detection of dark matter now pose a significant problem for
these scenarios. In addition, in order to avoid having a cold dark matter
density that is too large, such scenarios imply that the lightest
Higgs decays predominantly into inert neutralinos \cite{Hall:2010ix},
which is now ruled out by measurements of the Higgs couplings.  

A solution to this was already proposed in \cite{Hall:2011zq}, initially
motivated by trying to have a relic density compatible with
the cE$_6$SSM, where the $f^u$ and $f^d$ couplings vanish. To do this
one can use an exact $Z_2^S$ under which only the two inert-singlets,
$S_\alpha$, are odd. The inert-singlinos are then massless and
eventually contribute only a small amount to the effective number of
neutrinos. The dark matter candidate is then formed from the
neutralino sector which comprises of the bino, wino, Higgsinos and inert-Higgsinos. In such scenarios a bino-like dark matter candidate may
fit the measured relic density via a mechanism whereby the bino
scatters inelastically off SM states into heavier inert-Higgsinos. In this mechanism the $Z_2^H$ violating parameters that mix
the inert-Higgsinos with the other neutralinos play a vital role.

Thus, we actually have three discrete symmetries restricting the terms
in our superpotential. Such symmetries should be derived in an elegant
way from the high-scale physics. However, since there can be more than
one way to do this, leading to different couplings being suppressed,
we instead choose to take a more phenomenological approach. We
assume that only the $Z_2^S$ symmetry is exact, in order to avoid the severe problems
introduced by decays to light singlinos. On the other
  hand, not only is the $Z_2^H$ symmetry merely approximate, but some
  of the couplings which it is supposed to suppress could be quite
  large from a phenomenological point of view.  We therefore include
  $Z_2^H$ parameters that affect the neutralino masses in our
  analysis, performing scans involving $Z_2^H$ violating parameters
  for the first time.

  Since the E$_6$SSM is a broken supersymmetric model, like the MSSM, it has a
  large number of soft masses which parameterise the many ways that
  supersymmetry can be broken softly.  However here we will assume
  minimal supergravity inspired relations amongst the soft masses,
  which hold true at the gauge coupling unification scale where we
  assume there is an $E_6$ grand unified theory (GUT).  At this GUT
  scale we introduce a universal soft scalar mass ($m_0$) which all
  soft scalar masses are set equal, a universal gaugino mass,
  $M_{1/2}$, which all soft breaking gaugino masses are set equal to
  and a universal trilinear, $A_0$ which all the soft trilinears are
  set equal to. These universality conditions define the constrained
  version of the E$_6$SSM (cE$_6$SSM).
  
\subsection{Neutralino and chargino mass mixing matrices}

Our dark matter candidate is the lightest neutralino, $\tilde{\chi}^0_1$, which interacts with nucleons via spin-1 $Z$ exchange (spin-dependent), Higgs exchange (spin-independent) and squark exchange (both spin-dependent and spin-independent). It is not the lightest $R$-parity odd state, since there also exist massless inert-singlinos $\tilde{\sigma}$. Despite this, it is still stable and thus viable as a dark matter candidate, since it cannot decay to $\tilde{\sigma}$: the potential $\tilde{\chi}^0_1 \rightarrow \tilde{\sigma}\sigma$ decay has no kinematically viable final states with the same quantum numbers as the lightest neutralino \cite{Hall:2011zq}. We focus on the spin-independent component of the neutralino-hadron cross-section, since this is overwhelmingly dominant in most direct-detection experiments. 

The presence of additional fields lends a certain richness to the
content of the neutralino and chargino mass mixing matrices. If the $Z_2^H$ violating couplings in the E$_6$SSM are included, the lightest
neutralino may have as many as twelve contributing fields in its
interacting basis; if all $Z^H_2$ violating couplings are neglected,
however, this is reduced to six, since all interactions between third
and first/second generation Higgsinos are suppressed:

\begin{equation}
\tilde{N}_{int} = \left(
\begin{array}{cccc|cc}
\tilde{B} & \tilde{W} & \tilde{H}^0_d & \tilde{H}^0_u & \tilde{S} & \tilde{B}'
\end{array}
\right)^T.
\end{equation}
For this exploration of the EZSSM parameter space, these $Z^H_2$ violating couplings were allowed and we adhered instead to the exact $Z^S_2$ symmetry, resulting in a basis composed of ten fields ($\tilde{S}_u$ and $\tilde{S}_d$ are decoupled):
\begin{equation}
\tilde{N}_{int} = \left(
\begin{array}{cccccc|cccc}
\tilde{B} & \tilde{W}^3 & \tilde{H}^0_{d} & \tilde{H}^0_{u}  & \tilde{S}_3 & \tilde{B}' & \tilde{H}^0_{d1} & \tilde{H}^0_{d2} & \tilde{H}^0_{u1} & \tilde{H}^0_{u2} \\
\end{array}\right)
^T.
\end{equation}
This leads to the following neutralino mass mixing matrix:
\begin{equation}
M^N = \left(
\begin{footnotesize}
\begin{array}{cccccc|cccc}
M_1 & 0 & -\frac{1}{2}g'v_d & \frac{1}{2}g'v_u & 0 & 0 & 0 & 0 & 0 & 0 \\
0 & M_2 & \frac{1}{2} g v_d & -\frac{1}{2}gv_u & 0 & 0 & 0 & 0 & 0 & 0 \\
-\frac{1}{2}g'v_d & \frac{1}{2}gv_d & 0 & -\mu & -\frac{\lambda v_u}{\sqrt{2}} & Q_d g'_1v_d & 0 & 0 & -\frac{\lambda_{331}s}{\sqrt{2}} & -\frac{\lambda_{332}s}{\sqrt{2}} \\
\frac{1}{2}g'v_u & -\frac{1}{2}gv_u & -\mu & 0 & \frac{\lambda v_d}{\sqrt{2}} & Q_ug'_1v_u & -\frac{\lambda_{313}s}{\sqrt{2}} & -\frac{\lambda_{323}s}{\sqrt{2}} & 0 & 0 \\
0 & 0 & -\frac{\lambda v_u}{\sqrt{2}} & -\frac{\lambda v_d}{\sqrt{2}} & 0 & Q_s g'_1 s & -\frac{\lambda_{313}v_u}{\sqrt{2}} & -\frac{\lambda_{323}v_u}{\sqrt{2}} & -\frac{\lambda_{331}v_d}{\sqrt{2}} & -\frac{\lambda_{332}v_d}{\sqrt{2}} \\
0 & 0 & Q_dg'_1v_d & Q_ug'_1v_u & Q_sg'_1s & M'_1 & 0 & 0 & 0 & 0 \\
\hline
0 & 0 & 0 & -\frac{\lambda_{313}s}{\sqrt{2}} & -\frac{\lambda_{313}v_u}{\sqrt{2}} & 0 & 0 & 0 & -\frac{\lambda_{311}s}{\sqrt{2}} & -\frac{\lambda_{312}s}{\sqrt{2}} \\
0 & 0 & 0 & -\frac{\lambda_{323}s}{\sqrt{2}} & -\frac{\lambda_{323}v_u}{\sqrt{2}} & 0 & 0 & 0 & -\frac{\lambda_{321}s}{\sqrt{2}} & -\frac{\lambda_{322}s}{\sqrt{2}} \\
0 & 0 & -\frac{\lambda_{331}s}{\sqrt{2}} & 0 & -\frac{\lambda_{331}v_d}{\sqrt{2}} & 0 & -\frac{\lambda_{311}s}{\sqrt{2}} & -\frac{\lambda_{312}s}{\sqrt{2}} & 0 & 0 \\
0 & 0 & -\frac{\lambda_{332}s}{\sqrt{2}} & 0 & -\frac{\lambda_{332}v_d}{\sqrt{2}} & 0 & -\frac{\lambda_{321}s}{\sqrt{2}} & -\frac{\lambda_{322}s}{\sqrt{2}} & 0 & 0 \\
\end{array}
\end{footnotesize}
\right),
\end{equation}
where $Q_d = -\frac{3}{\sqrt{40}}$, $Q_u = -\frac{2}{\sqrt{40}}$ and $Q_s = -\frac{5}{\sqrt{40}}$ are the $U(1)_N$ charges of the down-type Higgs doublets, the up-type Higgs doublets and the SM-singlets respectively. Furthermore, $M_1$, $M_2$ and $M'_1$ are soft gaugino masses, while $g'_1$ is the GUT normalised $U(1)_N$ gauge coupling. The top-left block of this matrix is the usual NMSSM neutralino mass mixing matrix with an additional row and column for the $U(1)$ bino - this block will be referred to as the USSM sector. The rest are contributions from couplings with the inert-Higgsinos. Note that if the approximate $Z^H_2$ symmetry were to be enforced (by limiting $\lambda_{3\alpha3 }$ and $\lambda_{33\alpha}$ from above by imposing flavour changing neutral current (FCNC) constraints), the bottom right corner would become an approximately decoupled block diagonal mass matrix in a basis consisting of the inert-Higgsinos. For completion, we also write down the interaction basis of the chargino:
\begin{equation}
\tilde{C}_{int} = \left(
\begin{array}{cc|cc|cc|cc}
\tilde{W}^+ & \tilde{H}^+_{u3} & \tilde{H}^+_{u2} & \tilde{H}^+_{u1} & \tilde{W}^- & \tilde{H}^-_{d3} & \tilde{H}^-_{d2} & \tilde{H}^-_{d1} \\
\end{array}
\right)^T.
\end{equation}
The chargino mass mixing matrix is:
\begin{equation}
M^C = \left(
\begin{array}{cc}
0 & P^T \\
P & 0 \\
\end{array}
\right),
\end{equation}
where
\begin{equation}
P = \left(
\begin{array}{cc|cc}
M_2 & \sqrt{2}m_Ws_\beta & 0 & 0 \\
\sqrt{2}m_Wc_\beta & \mu & \frac{1}{\sqrt{2}}\lambda_{332}s & \frac{1}{\sqrt{2}}\lambda_{331}s \\
\hline
0 & \frac{1}{\sqrt{2}}\lambda_{323}s & \frac{1}{2}\lambda_{322}s & \frac{1}{\sqrt{2}}\lambda_{321}s \\
0 & \frac{1}{\sqrt{2}}\lambda_{313}s & \frac{1}{\sqrt{2}}\lambda_{312}s & \frac{1}{\sqrt{2}}\lambda_{311}s \\
\end{array}
\right).
\end{equation}
\section{Scan procedure}
Following the considerations in the previous section, we perform a
scan over $x_{u1}$, $x_{d1}$, $x_{u2}$, $x_{d2}$, $\lambda_{11}$,
$\tan\beta$, $\lambda$, $\lambda_{22}$, $s$ and $\kappa$, where
$x_{u1}$ and $x_{u2}$ are the $S H_{dj} H_{u3}$ couplings with $j = 1,
2$, $x_{d1}$ and $x_{d2}$ are the $S H_{d3} H_{uk}$ couplings with $k
= 1, 2$, and finally $\lambda_{mn}$ are the $S H_{dm} H_{un}$
couplings with $m, n = 1, 2$.  We do not scan over the universal soft
masses as these are output parameters determined by the spectrum
generator in our setup, as will be explained shortly.

The large dimensionality of this
parameter set makes a random or grid scanning method prohibitively
expensive.  For efficient sampling we use {\tt Multinest-2.4.5} which
employs a nested sampling algorithm to calculate the Bayesian evidence
of the model by Monte Carlo integration, obtaining posterior samples
as a by-product~\cite{MultiNest,MultiNest2,MultiNest3}.

%

However in this study we do not consider the Bayesian evidence or the
posterior samples. Instead, we simply use {\tt Multinest} as a tool to quickly
find E$_6$SSM parameters that give rise to the observed relic
abundance of dark matter whilst remaining consistent with the LHC
Higgs mass measurement, and have a WIMP-nucleon cross-section for the
lightest neutralino that lies close to the current experimental
exclusion limits.  To do this, we passed the following `likelihood'
function to
Multinest:
\begin{equation}
\textrm{log} L = -\left(\frac{m_{h_1} - m^{\text{ex}}_{h_1}}{\sigma^{m_{h_1}}}\right)^2 -\left(\frac{\Omega h^2 - (\Omega h^2)^{\text{obs}}}{\sigma^{\Omega h^2}}\right)^2 - \left(\frac{\sigma_{\text{SI}} - \sigma_{\text{SI}}^{\textrm{lim}}}{0.5\sigma_{\text{SI}}^{\textrm{lim}}}\right)^2.
\label{loglike}
\end{equation}

Note that the density of points in our final plots will not have a clear meaning, and we will instead only focus on the type of dark matter solution that we encounter.

The first term is the constraint from the LHC Higgs mass measurement,
where we use the 2012 CMS result $m_h^{\text{ex}} = 125.3$ GeV with
$\sigma^{m_{h_1}} = 0.64$ GeV, consisting of a quadrature sum of the
quoted systematic and statistical errors
\cite{2012PhLB..716...30C}. This has since been improved to
$m_h^{\text{ex}} = 125.09$ GeV with $\sigma^{m_{h_1}}=0.24$ GeV by combined
CMS and ATLAS measurements \cite{2015arXiv150307589A}, but the details
will not affect our final conclusions. The second term is the
constraint from the relic density, assuming a central value of
$(\Omega h^2)^{\text{obs}} = 0.1196$. This is using the 2013 value
from the Planck collaboration along with associated uncertainty
$\sigma^{\Omega h^2} = 0.0031$\cite{PlanckCollab}. The third and final
term is the constraint on the WIMP-nucleon SI cross-section from the
2013 LUX results \cite{Akerib:2013tjd}. Here, the function
$\sigma_{\text{SI}}^{\textrm{lim}}$ was extrapolated from the 95\%
confidence level LUX limit, and the final term ensures that we find
solutions close to the current experimental reach.  The LUX results
have been updated recently \cite{Akerib:2016vxi} imposing
substantially stronger limits on the spin-independent cross-section.
Nonetheless we also compare our final results from the scan with these
recent LUX results which appeared after the scan had completed. The
large width of the Gaussian function used above is sufficient to give
us solutions that are beyond the current LUX reach, however, and we
briefly comment on the impact in the next section. We assume a flat
prior on all parameters, and scan within the ranges given in
Table~\ref{parameters}.

\begin{table}[h!!]
  \begin{center}
    \begin{tabular}{| c | c || c | c |}
      \hline
      Parameter & Range & Parameter & Range \\
      \hline
      \hline
      $x_{u1}$ & $0 - 0.5$ & $x_{u2}$ & $0 - 0.5$ \\
      \hline
      $x_{d1}$ & $0 - 0.5$ & $x_{d2}$ & $0 - 0.5$ \\
      \hline
      $\lambda_{11}$ & $0.0001 - 1.0$ & $\lambda_{22}$ & $0.0001 - 1.0$ \\
      \hline
      tan$\beta$ & 1 - 40 & $s$ & $0 - 100000$ \\
      \hline
      $\lambda$ & $-0.5 - 0.5$ & $\kappa$ & $0 - 5$ \\
      \hline
    \end{tabular}
   \caption{The parameters used in our scan, along with the allowed ranges.}
   \label{parameters}
  \end{center}
\end{table}


For each point in our scan, we calculate the mass spectrum using an unpublished spectrum generator that uses semi-analytic solutions for the soft masses as described in Refs.\cite{cE6SSM1,cE6SSM2}.  The semi-analytic solutions express the soft masses, including those appearing in the electroweak symmetry breaking conditions, in terms of the universal soft masses, $m_0$, $M_{1/2}$ and $A_0$ which are fixed at the GUT scale. As a result the universal softmasses can be parameters which are fixed by the electroweak symmetry breaking conditions. Without this procedure it is hard to solve the constrained version of the model, as one wants to fufill the EWSB constraint by fixing a softmass at the electroweak scale, while also requiring it fulfills the high scale universality condition.  We run all soft masses, superpotential parameters and gauge couplings between the electroweak and GUT scale with the full two-loop RGEs, by linking to {\tt FlexibleSUSY}~\cite{FlexibleSUSY,Athron:2014wta}, which uses  {\tt SARAH} \cite{SARAH,Staub:2010jh,Staub:2009bi,Staub:2012pb,Staub:2013tta} and numerical routines from {\tt SOFTSUSY} \cite{SOFTSUSY,Allanach:2013kza}. The Higgs mass is calculated by generalising an NMSSM calculation using EFT techniques but expanded to fixed two-loop order, as described in \cite{King:2005jy,cE6SSM2}.  Since we expect at the outset to have a very heavy SUSY scale and also allow exotic couplings in the scan to be large, using the full two-loop fixed order calculation  for this model \footnote {Recently this has been made possible with {\tt SARAH} / {\tt SPheno} \cite{Goodsell:2014bna}.} or an MSSM effective field theory computation \footnote{As was done in Ref.\cite{Athron:2016gor} using {\tt SUSYHD}\cite{Vega:2015fna}.} would not significantly improve the precision, while an E$_6$SSM effective field theory computation was not available when this work was performed \footnote{Such a calculation \cite{Athron:2016fuq} was made available while this paper was being finalised.}.   We do not expect our results to be substantially changed by a more accurate determination of the Higgs mass. The relic density of dark matter and WIMP-nucleon cross-section for the lightest neutralino are obtained using a version of {\tt micrOMEGAs-2.4.5} \cite{Belanger:2006is,Belanger:2008sj, Belanger:2010pz}, which was extended\footnote{We thank Jonathan Hall for supplying us with this version of {\tt micrOMEGAs}.} for the E$_6$SSM with an E$_6$SSM {\tt CalcHEP} \cite{Belyaev:2012qa} model file generated using {\tt LanHEP} \cite{Semenov:2010qt}. 

As well as using the measured Higgs mass, dark matter relic density and LUX limits to guide the scan we also apply explicit experimental constraints to the data before plotting results.  Unless explicitly stated otherwise we require that each point fulfills the following:
\begin{eqnarray}
&&(\Omega h^2)^{\text{obs}} - 2 \sigma^{\Omega h^2} > \Omega h^2 > (\Omega h^2)^{\text{obs}} + 2 \sigma^{\Omega h^2} \label{Eq:OmegaRange}\\
  && 122.3 \, \text{GeV} < m_h < 128.3 \, \text{GeV} \label{Eq:MhRange}\\
  && m_{\text{gluino}} > 1.4 \, \text{TeV} \label{Eq:MgluinoRange}\\
  && M_{Z^\prime} > 2.85 \, \text{TeV} \label{Eq:MZprimeRange}\\
  && \mu_{D_i} > 1.4 \, \text{TeV} \label{Eq:DiquarkRange}\\
  && m_{\chi_i^\pm} > 100 \, \text{GeV} \label{Eq:CharginoRange}
\end{eqnarray}
Here we give a large $6$ GeV range for the Higgs mass, $m_h$ to
account for the well known large theoretical errors associated with
this prediction. Since the scan was designed to efficiently find
points with a Higgs mass prediction close to the experimentally
measured value this does not cut out many points. The constraint on
the relic density ensures that we can explain all of the dark matter
relic abundance, while not over closing the universe.  Since the focus
of our work is the direct detection phenomenology of the E$_6$SSM, we
do not include collider constraints in our scan.  However, as has been
discussed previously \cite{cE6SSM4}, the hierarchical spectrum in the
constrained E$_6$SSM means that sfermions will be safe from LHC limits
so long as the gluino is above the CMSSM limit in the heavy sparticle
limit, which is what we impose here\footnote{In cases where there are
  additional light neutralinos compared to the MSSM the gluino cascade
  decay can be modified, which can alter the gluino mass limit
  \cite{Belyaev:2012si,Belyaev:2012jz}.  However this would not have a
  large impact on our results.}.  We also require that the exotic
coloured fermions, which could potentially be light, have a mass,
$\mu_{D_i}$, greater than $1.4$ TeV, since we expect the signature to
be comparable to that of the gluinos, though no dedicated quantitative
analysis has been done for these states. At the same time, LEP limits
on charginos should be rather robust and we use these to set a lower
limit on the lightest chargino states in this model.  Finally we use
the latest $Z^\prime$ limits to ensure that this would not have been
discovered as a peak in the dilepton invariant mass spectrum at the
LHC.

\section{Results and discussion}
\subsection{Dark Matter candidates and their spin-independent cross-section}

\begin{figure}[h!!]
  \begin{center}
    \includegraphics[width=0.49\textwidth]{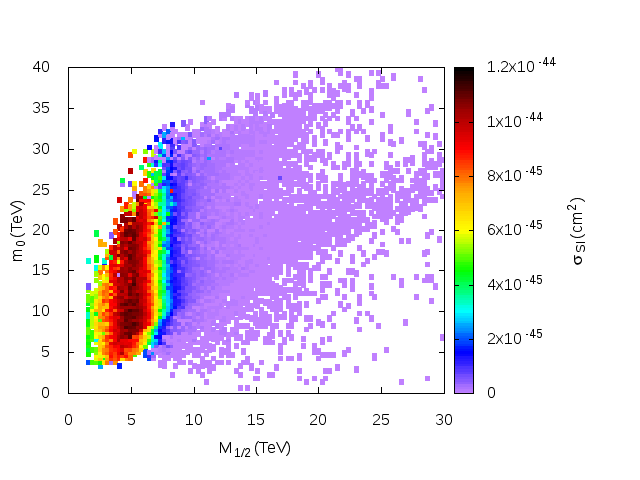}%
     \includegraphics[width=0.49\textwidth]{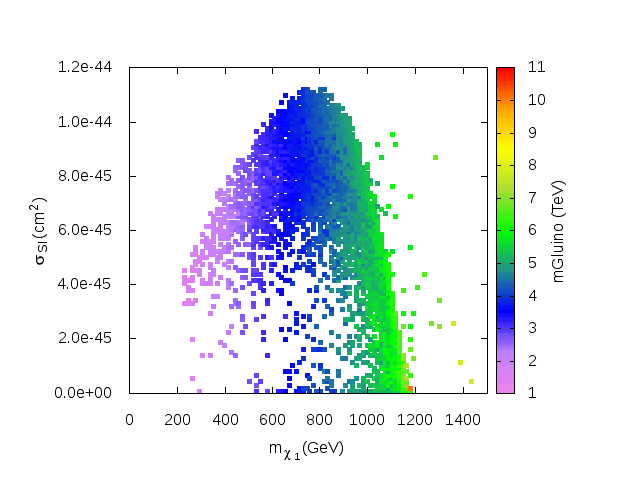}
  \caption{The spin-independent cross-section for direct detection of
    dark matter for all points found in the scan consistent with
    Eqs.~\ref{Eq:OmegaRange}--\ref{Eq:CharginoRange}.  Left panel: the
    $m_0-M_{1/2}$ plane with the maximum binned value of $\sigma_{SI}$
    given as the colour contour. Right panel: $\sigma_{SI}$ against
    the lightest neutralino mass, $m_{\chi_1}$. A colour contour of
    the lightest gluino mass in each bin is shown to indicate in
    which cases gluino production could be observed at the
    LHC. }
\label{fig:generalresults}%
\end{center}
\end{figure}

We now turn to a discussion of the results of our scan, including the
possible dark matter explanations that have been revealed and the
implications for these from dark matter direct detection experiments.
The spin-independent cross-section ($\sigma_{SI}$) for direct
detection of dark matter is shown in Figure \ref{fig:generalresults}
for all points which pass our experimental constraints given in
Eqs.~~\ref{Eq:OmegaRange}--\ref{Eq:CharginoRange}. In the left panel
$\sigma_{SI}$ is shown as a colour contour in the $m_0-M_{1/2}$
plane. Care should be taken when interpreting this plot as the very
different renormalisation group flow of the E$_6$SSM, compared to the
MSSM, means that the relationship between soft masses at low energies
(and thereby the physical mass eigenstates) and universal soft masses
at the GUT scale is changed considerably. The right panel shows
$\sigma_{SI}$ plotted directly against the neutralino mass, with the
minimum gluino mass from each bin plotted as a colour contour.

The left panel shows that we can explain the full relic abundance of
dark matter, while satisfying collider constraints, for much of the
$m_0-M_{1/2}$ plane. Comparing this to the right panel we see that
this happens for dark matter candidates with a wide range of masses,
though the density of solutions found varies a lot.  The correct relic
density is achieved through several different mechanisms, which depend on
the nature of the dark matter candidate. In this model the dark matter candidate is the lightest neutralino, which may be bino-like, Higgsino-like, inert-Higgsino-like or some combination of two or all three of these; we will discuss each case briefly.

As can be seen in the left panel many solutions we have found go way
beyond the reach of the LHC.  While the heavy SUSY scale there makes
it very challenging to predict the Higgs mass precisely we expect that
our result here should be reproducible with higher precision
calculations that have been recently developed \cite{Athron:2016fuq},
requiring only adjustments to parameters that are essentially orthogonal to
the other predictions we present.  When $M_{1/2} \gtrsim 8$ TeV this
implies that $M_1$ is significantly larger than $\approx 1$ TeV and
the correct relic density can be explained without a large bino
component to the dark matter. As the colour contour in the right panel
of Figure \ref{fig:generalresults} shows, the large $M_{1/2}$ values
required for these points means that the gluino is very heavy and well
beyond the reach of the LHC. In this case the dark matter candidate
is either pure Higgsino, pure inert-Higgsino or a mixture of the two
and these solutions are found in the dense almost vertical band of
solutions shown in the lower right region of the right panel of Figure
\ref{fig:generalresults}.

\begin{figure}[h!!]
  \begin{center}
    \includegraphics[width=0.49\textwidth]{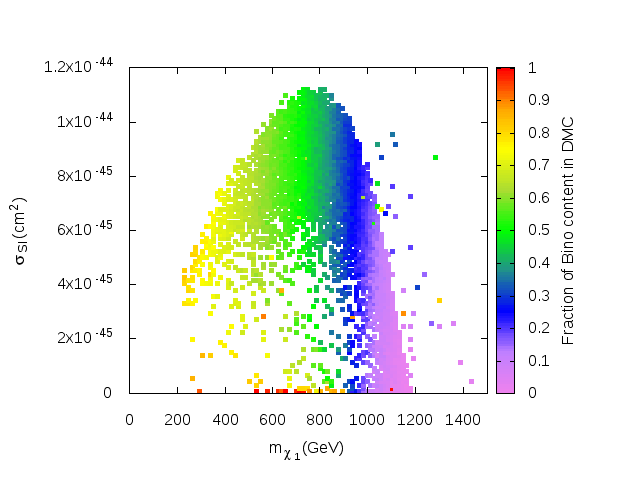}
    \includegraphics[width=0.49\textwidth]{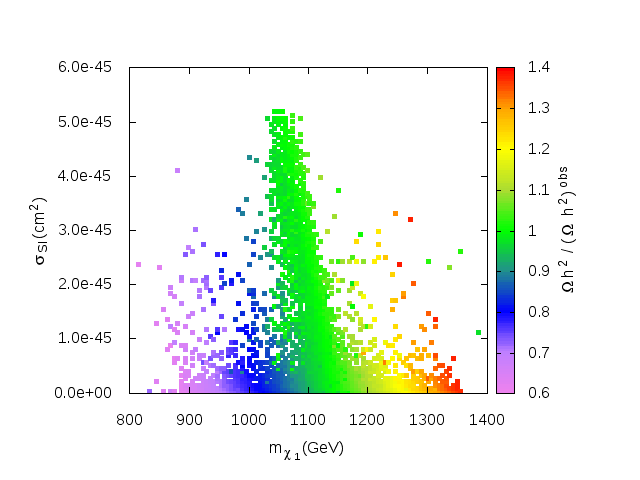}   
    \caption{Spin-independent cross-section, $\sigma_{SI}$, against the lightest neutralino mass, $m_{\chi_1}$.  In the left panel we show the bino content of the dark matter as a colour contour and plots all points found in the scan that satisfy the experimental constraints in Eqs.~\ref{Eq:OmegaRange}--\ref{Eq:CharginoRange}.  The right panel shows a colour contour of the minimum value of $\Omega h^2 / (\Omega h^2)^{obs}$, in each bin for points with less than 10\% bino content that satisfy all constraints in Eq.~\ref{Eq:MhRange}--\ref{Eq:CharginoRange}, omitting only the condition on the relic density so that the variation can be shown. }
\label{fig:HiggsinoIHDM}%
\end{center}
\end{figure}

This is confirmed in Figure \ref{fig:HiggsinoIHDM} where in the left
panel the bino content is shown varying across the
$m_{\chi_1^0}-\sigma_{SI}$ plane. The Higgsino and inert-Higgsino dark
matter candidates both obtain the correct relic density through the
same annihilation mechanisms as Higgsino dark matter in the MSSM, which
is why these scenarios have a mass of around $1$ TeV.

For dark matter candidates with no bino content (defined here as
having less than 10\% bino component) the standard co-annihilation
mechanism does not allow the correct relic density to be obtained
outside of this band.  When such a dark matter candidate is lighter
than this it will typically give a relic density which is too large as
a light Higgsino annihilates too efficiently, as can be seen in the
right panel of Figure \ref{fig:HiggsinoIHDM}. Similarly if the mass is
larger than the masses in this band then the dark matter will
not annihilate enough leading to over-closure of the universe.

This prediction can be evaded if there is a non-standard mechanism for
Higgsino dark matter, such as a funnel region. Indeed the small number of
scattered Higgsino or inert-Higgsino points that still fit the relic
density very well, while having $m_{\chi_1^0} > 1.2$ TeV, correspond
to A-funnel scenarios where the pseudoscalar Higgs mass is very close
to being twice the mass of the lightest neutralino.

Another possibility is that the neutralino is predominantly bino.
Pure bino scenarios have very low spin-independent direct detection
cross-sections, as the SM-like Higgs exchange diagram is suppressed.
However in the MSSM, the current mass limits on sparticles make it quite
difficult to successfully achieve the correct relic density for a pure
bino.  In contrast in the cEZSSM there is a special mechanism which can
achieve the correct relic density with a predominantly bino-like dark
matter candidate that was proposed in Ref.~\cite{Hall:2011zq}. There
the relic density is achieved in a manner which is not possible in the
MSSM, involving scattering off of standard model states into
inert-Higgsinos, which must not be much heavier than the bino.

In addition we also find scenarios where the dark matter candidate is
pure bino and the mass is around half that of the pseudoscalar Higgs
boson.  This gives us the bino A-funnel scenario, where, as in the MSSM,
this tuning allows the annihilation cross-section to be large enough
that the pure bino candidate does not over-close the universe.  

Another possibility to obtain the measured relic density with a
lightest neutralino mass lower than $\approx 1$ TeV, away from this
Higgsino/inert-Higgsino band, is to tune the parameters to lie in the
well-tempered region \cite{ArkaniHamed:2006mb}.  As in the CMSSM the
wino is always heavier than the bino, so such scenarios will have a
significant admixture of bino and inert-Higgsino or Higgsino dark
matter, as can be seen in the left panel of Figure
\ref{fig:HiggsinoIHDM}.  While scenarios where the dark matter candidate is
predominantly composed of just one gauge eigenstate have a suppressed
spin-independent cross-section (and in the case of Higgsinos and inert-Higgsinos a very heavy mass spectrum), scenarios with mixed dark matter
candidates can be quite different.

\begin{figure}[h!!]
  \begin{center}
    \includegraphics[width=0.49\textwidth]{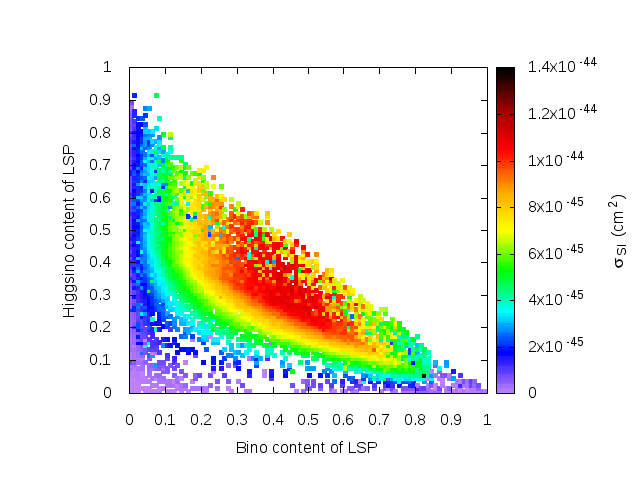}%
    \includegraphics[width=0.49\textwidth]{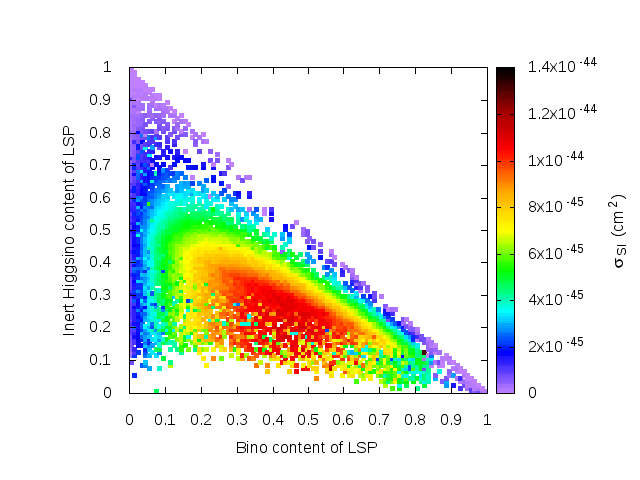}\\
\includegraphics[width=0.49\textwidth]{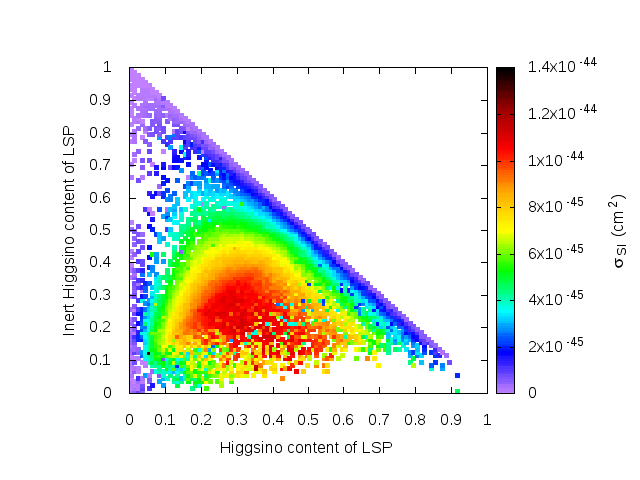}
\caption{Spin-independent cross-section, $\sigma_{SI}$, varying with the content of the lightest neutralino. To illustrate the mechanism clearly, we do not impose any experimental constraints in these plots.  In the top left panel we plot $\sigma_{SI}$ as a colour contour with the bino content on the x-axis and the Higgsino content on the y-axis.  In the top right panel we show the same, but with the inert-Higgsino content on the y-axis instead of the Higgsino content.  In the bottom panel we have Higgsino content on the x-axis and inert-Higgsino content on the y-axis.}
\label{fig:AdmixVsSICS}%
\end{center}
\end{figure}

In particular, it is well known that in the MSSM one may also
obtain the correct relic density for mixed bino-Higgsino candidates
\cite{ArkaniHamed:2006mb}.  This scenario avoids requiring $M_1$ to be
significantly greater than $m_{\chi^0_1}\approx 1$ TeV, and therefore
gives rise to better prospects for discovery in collider experiments.
However, introducing more bino-Higgsino mixing enhances the
direct search cross-section by increasing the contribution from Higgs exchange, as
can be seen in the top left panel of Figure \ref{fig:AdmixVsSICS}, where we plot
$\sigma_{SI}$ as a colour contour with the Higgsino, and bino content
as the axes.  This has already been discussed in
Ref.~\cite{Athron:2016gor} for constrained versions
of the MSSM and an alternative $E_6$-inspired model, where it was
shown that the bino-Higgsino mixing is now heavily constrained by
LUX \cite{Akerib:2013tjd,Akerib:2015rjg,Akerib:2016vxi}.

However, unlike the $E_6$-inspired models considered in
Refs.\cite{Athron:2015vxg, Athron:2016gor}, in the cEZSSM there are
further possibilities involving the inert-Higgsinos. Since associated
inert-Higgs bosons are all very heavy, the s-channel annihilation
diagram involving the inert-Higgs is suppressed and in this case the
correct relic density is simply obtained by diluting the
inert-Higgsino co-annihilation mechanism through the reduced
inert-Higgsino content. The heavy inert-Higgs states also mean that
the direct detection cross-section is suppressed as is shown in the top
right panel of Figure \ref{fig:AdmixVsSICS}, so the inert-Higgsinos
mixing with the bino does not lead to large spin-independent
cross-sections.  There is no bino-Higgs-inert-Higgsino for a SM-like
Higgs exchange contribution and the inert-Higgs scalar is very heavy,
which suppresses an inert-Higgs exchange contribution to the
cross-section. Note that these plots include scenarios where the dark
matter candidate contains significant admixtures of all three types
(Higgsino, inert-Higgsino and bino) of gauge states. This is why the
cross-section can become large here as well for moderate values of the
bino and inert-Higgsino contents, where they do not sum to unity.


\subsection{Impact of direct detection experiments}
\begin{figure}[h!!]
  \begin{center}
    \includegraphics[width=0.7\textwidth]{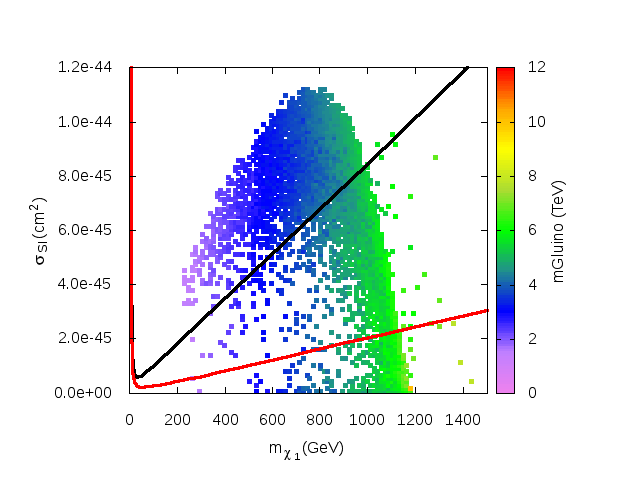}
    \caption{Spin-independent cross-section, $\sigma_{SI}$, against the lightest neutralino mass, $m_{\chi_1}$ with the minimum gluino mass in each bin plotted as a colour contour. All experimental constraints from Eqs.\ref{Eq:OmegaRange}--\ref{Eq:CharginoRange} are applied. The black curve shows the LUX 2015 limit \cite{Akerib:2015rjg}, while the LUX 2016 limit \cite{Akerib:2016vxi} is indicated by the red curve.   }
\label{fig:LUXvsOurScan}%
\end{center}
\end{figure}

 Applying the LUX 2015 and LUX 2016 constraints to our results, as shown
 in Figure \ref{fig:LUXvsOurScan} demonstrates the dramatic impact of
 LUX 2016, ruling out many scenarios. As one could anticipate the pure
 Higgsino/inert-Higgsino scenarios can survive, and these correspond
 to the large region of $m_0-M_{1/2}$ parameter space at larger
 $M_{1/2}$ where the spin-independent cross-section is rather
 small. Note that in this case the limit on $M_{1/2}$ for Higgsino
 dark matter set by the LUX experiment exceeds the LHC reach considerably.
 However scenarios with a sub-TeV dark matter candidate can be more
 relevant to collider phenomenology.

Since the scan was designed to find scenarios close to the direct
detection cross-sections limits of LUX 2013, it is not surprising that
so many scenarios we found are now ruled out.  Nonetheless the results
have still revealed the possibility of mixed bino inert-Higgsino dark
matter, and in these cases the cross-section can be considerably
weaker, while still fitting the relic density.

While only a small number of these points lying below the LUX 2016
limit were found, they do provide a novel way to escape the stringent
limits from the latest direct detection experiments.

To illustrate the scenarios which can evade the LUX limits we present
five benchmark scenarios in Table \ref{benchmarksfirst}.  These benchmark scenarios
represent the different mechanisms we found where the relic density
can be fitted while still evading the LUX limits.  These possibilities
are as follows.

First the dark matter may simply be composed of only Higgsino or inert-Higgsino gauge states.  Higgsino dark matter is a well known
possibility in the MSSM and has also been studied in other $E_6$-inspired scenarios that have been explored previously
\cite{Athron:2016gor}.  BM1 is a scenario where the dark matter relic
density is explained from a neutralino dark matter candidate that is
predominantly inert-Higgsino in nature.  The dominant channels in this
case, are chargino-neutralino co-annihilations, as is the case for
standard Higgsino dark matter.  Since this inert-Higgsino dark matter
candidate has a mass of $1.1$ TeV, the observed relic density can be
fitted, while the spin-independent cross-section is sufficiently small
that the LUX limits can be evaded.

Typically if a pure Higgsino or inert-Higgs dark matter candidate has a mass much lower than that of BM1 the predicted relic density will be too small, while if the mass is much higher then it will be too large, leading to over-closure of the universe.  However the latter can be avoided if these annihilations are enhanced by a funnel mechanism.   BM2 shows a Higgsino
dark matter candidate where the pseudoscalar Higgs boson has a mass,
$m_{A^0} \approx 2 m_{\chi_1^0}$, allowing the observed relic density
to be achieved predominantly through near-resonant annihilation through
the pseudoscalar Higgs boson into $b\overline{b}$.  Such scenarios are commonly
referred to as A-funnel scenarios in the literature.

For lighter dark matter, one may consider special bino dark matter
scenarios.  BM3 shows a dark matter candidate which is made up
primarily of the bino gauge eigenstate.  The relic density for
these scenarios is satisfied through the mechanism previously explored
in Ref.~\cite{Hall:2011zq}, which requires that there is a
predominantly inert-Higgsino neutralino with a mass very close to the
lightest neutralino.  This mechanism proceeds by the lightest
neutralino up-scattering into the slightly heavier inert-Higgsino
neutralino, which then co-annihilates at a rate large enough to fit the
observed relic density.  BM4 shows another possibility where the
bino-like dark matter candidate annihilates through the pseudoscalar
Higgs boson into mostly $b\overline{b}$, giving another A-funnel
possibility, a type of scenario that is well known in the MSSM and has
also been looked at in $E_6$-inspired models previously \cite{Athron:2016gor}.

Finally BM5 shows a new possibility that has not been discussed
previously in the literature.  In this scenario the dark
matter candidate has large admixtures of bino and
inert-Higgsino.  While scenarios where the bino mixes only with a
Higgsino are heavily constrained due to the large spin-independent
cross-section obtained through Higgs exchange, these scenarios are
free of this problem since there is no light inert-Higgs state to
give rise to a large inert-Higgs exchange contribution to the spin-independent cross-section. Similarly unlike standard cases of a mixed
Higgsino-bino candidate where the relic density is achieved by
annihilation mostly through the light Higgs boson\footnote{See for
  example benchmarks given in Ref.~\cite{Athron:2015vxg}.}, in this
case the observed relic density is achieved only through the usual
co-annihilation channels of Higgsino dark matter.  Over-closure of the
universe is avoided because of the bino mixing, which dilutes the
efficiency of this process.


%
\subsection{Conclusions and Outlook}
We have presented the most extensive phenomenological exploration of
the cE$_6$SSM to date, and revealed a large volume of parameter space
compatible with the measured relic abundance of dark matter and the
latest results from the LHC, including a 125 GeV Higgs boson and
collider limits on new states.  This work has revealed a number of
different scenarios for explaining the observed relic density of dark
matter. We have shown the significant impact of the recent direct detection limits.  However even with these tough limits there are
a number of mechanisms for obtaining the measured relic density that
can have a spin-independent direct detection cross-section below the
LUX 2016 limit.

In particular if the dark matter candidate is a mixture of
inert-Higgsino and bino it can be significantly lighter than 1 TeV
and still predict the correct relic density and evade the LUX 2016
limit for direct detection.  Another possibility in this model is a
pure bino dark matter candidate, where the relic density can be obtained either through an up-scattering into inert-Higgsinos which then co-annihilate with charged inert-Higgsinos, or through A-funnel
scenarios.  Such scenarios are more likely to be observed in the last part of LHC
Run II, or during subsequent runs at high luminosity. Certainly they have much better prospects for
  observability in collider experiments than the pure Higgsino or
  inert-Higgsino scenarios, where the gluino must be heavier than
  about $6$~TeV.

Nonetheless even the pure Higgsino and inert-Higgsino scenarios which we
explored here will be within range of XENON1T \cite{Xenon1T}.  The
XENON1T experiment is the third phase of the XENON experiment at the
Gran Sasso Laboratory and will soon begin to publish results. The
sensitivity of this experiment is expected to reach a minimum
spin-independent WIMP-nucleon cross-section of $1.6 \times 10^{-47}$
cm$^2$ at $m_{\chi} = 50$ GeV, a factor of approximately 50 times
better than the current LUX limit at the same WIMP mass
\cite{Xenon1T}.  This is sensitive enough to be able detect all of the
benchmark points we have presented and will provide severe constraints
on the model. 

\begin{table}[h!!]
\begin{center}
\footnotesize
\begin{tabular}{|c|c|c|c|c|c|}
\hline
& {\bf BM1} & {\bf BM2} & {\bf BM3} & {\bf BM4} & {\bf BM5}  \\
\hline
\hline
$\lambda$ & -0.0655 & 0.18122 & -0.552579 & 0.0831877  & 0.285842 \\
$\kappa$ & 0.2211 & 0.169603 & 0.22825 & 0.122173 & 0.230147  \\
tan$\beta$ & 29.3 & 22.4239 & 5.4816 & 38.5743 & 7.1522   \\
$s$ (GeV) & 52966.4 & 70523.5 & 16739.2 & 29474.3 & 73442.8   \\
$\lambda_{11}$  & 0.2435763862  & 0.316856  & 0.048464  & 0.0903254 & 0.617126   \\
$\lambda_{22}$  & 0.02779747502 & 0.654349  & 0.588343  & 0.795588 & 0.0171   \\
$x_{u1}$  & 0.1196894578  & 0.271995  & 0.0626005  & 0.0344383 & 0.167725   \\
$x_{d1}$  & 0.3668858675  & 0.169106  & 0.062843  & 0.464061  & 0.0910559   \\
$x_{u2}$  & 0.01118730707 & 0.0093813   & 0.434091  & 0.0691506  & 0.0179128   \\
$x_{d2}$  & 0.08349296723 & 0.439901  & 0.351041  & 0.270508 & 0.0641696   \\
\hline
$m_0$ (GeV) & 19262.0 & 26562.8 & 4069.08 & 13741.2 & 19330.3   \\
$M_{1/2}$ (GeV) & 7387.5  & 9492.72 & 4008.4 & 3467.43 & 3338.41   \\
$A_0$ & 6269.1 & 16767.1 &-574.161  & 10650.9 & 24157.8   \\
$\Omega h^2$ & 0.1190  & 0.1162 & 0.1180 & 0.1240 & 0.1223   \\
$\sigma_{SI}$ (cm$^2$) & 1.20 $\times 10^{-46}$  & 1.106 $\times 10^{-46}$ & 1.60 $\times 10^{-47}$ & 2.86 $\times 10^{-45}$ & 4.53 $\times 10^{-46}$   \\
\hline
$m_{\tilde{\chi}^0_1}$ (GeV) & 1104.0  & 1387.9 & 631.3 & 557.1 & 543.7   \\
$m_{\tilde{\chi}^0_2}$ (GeV) & 1106.0  & 1393.9 & 643.0 & 656.8 & 563.3   \\
$m_{\tilde{\chi}^0_3}$ (GeV) & 1167.9  & 1521.8 & 643.8 & 658.3 & 567.3   \\
$m_{\tilde{\chi}^0_4}$ (GeV) & 2069.4  & 2700.2 & 1118.8 & 1023.5 & 980.9   \\
$m_{\tilde{\chi}^0_5}$ (GeV) & 5300.8  & 21956.5 & 5621.4 & 9152.0 & 10496.1   \\
$m_{\tilde{\chi}^\pm_1}$ (GeV) & 1105.7  & 1392.3 & 642.9 & 648.9 & 562.5   \\
$m_{\tilde{\chi}^\pm_2}$ (GeV) & 2069.4  & 2700.2 & 643.0 &1023.4  & 980.9   \\
$m_{\tilde{\chi}^\pm_3}$ (GeV) & 5301.5  & 21956.6 & 5622.7 & 9152.0 & 10496.3   \\
$m_{h_1}$ (GeV) & 124.8  & 125.4 & 122.7 & 125.0 & 127.2   \\
$m_{A^0}$ (GeV) & 11900  & 2838.6 & 6329.1 & 1093 & 9393   \\
$m_{\tilde{t}_1}$(GeV) & 15200 & 19600 & 4920 & 9290 & 13300 \\
$m_{Z'}$ (GeV) & 19600 & 26100 & 6190 & 10905 & 27200  \\
\hline
$|Z(N)_{11}|^2$ & 0.0238  & 0.0292 & 0.924 & 0.901 & 0.788   \\
$|Z(N)_{12}|^2$ & 0.0003  & 0.000989 & 1.06 $\times 10^{-5}$& 0.00125 & 0.000354   \\
$|Z(N)_{13}|^2$ & 6.46 $\times 10^{-6}$  & 0.292 & 0.0001522 & 0.00211 & 0.00144  \\
$|Z(N)_{14}|^2$ & 0.0497  & 0.289 & 0.0003186 & 0.0407 & 0.00961  \\
$|Z(N)_{15}|^2$ & 1.72 $\times 10^{-7}$  & 6.41 $\times 10^{-7}$ & 7.03 $\times 10^{-9}$ & 4.61 $\times 10^{-7}$ & 1.43$\times 10^{-8}$    \\
$|Z(N)_{16}|^2$ & 1.36 $\times 10^{-9}$  & 3.96 $\times 10^{-7}$ & 1.15 $\times 10^{-8}$ & 7.27 $\times 10^{-9}$ & 7.21 $\times 10^{-11}$   \\
$|Z(N)_{17}|^2$ & 0.4886  & 0.132 & 7.50 $\times 10^{-5}$ & 0.000226 & 0.108   \\
$|Z(N)_{18}|^2$ & 0.4250  & 6.12 $\times 10^{-5}$ & 0.000196 & 0.000290 & 0.0921   \\
$|Z(N)_{19}|^2$ & 9.80 $\times 10^{-5}$  & 0.0624 & 0.0383 & 0.0546 & 3.60 $\times 10^{-5}$  \\
$|Z(N)_{110}|^2$ & 0.0123  & 0.193 & 0.0370 & 5.91 $\times 10^{-5}$ & 0.000713   \\
\hline
\end{tabular}
\caption{The five benchmark points chosen in this study. {\bf BM1} features a lightest neutralino with a high inert-Higgsino content. {\bf BM2} features a lightest neutralino that is a mixture of Higgsino and inert-Higgsino. The lightest neutralino of {\bf BM3} has a pure-bino character and the model satisfies the relic density constraint through the upscattering mechanism. {\bf BM4} also has a pure-bino LSP, and in this case the model achieves the correct relic density through resonant annihilation via the $A$ boson. Finally, {\bf BM5} has an LSP with a mixture of bino and inert-Higgsino components.}
\label{benchmarksfirst}
\end{center}
\end{table}

\begin{table}[h!!]
\begin{center}
\tiny
\begin{tabular}{|c|c|c|c|c|c|}
\hline
& {\bf BM1} & {\bf BM2} & {\bf BM3} & {\bf BM4} & {\bf BM5} \\
\hline
$\tilde{\chi}^0_1 \tilde{\chi}^+_1 \rightarrow t\bar{b}$ & 7.2\% & 12\% & 4\% & $<$1\% & 5\% \\
$\tilde{\chi}^0_1 \tilde{\chi}^+_1 \rightarrow u\bar{d}$ & 7.0\% & 5\% & 4\% & $<$1\% & 5\% \\
$\tilde{\chi}^0_1 \tilde{\chi}^+_1 \rightarrow c\bar{s}$ & 6.9\% & 5\% & 4\% & $<$1\% & 5\% \\
$\tilde{\chi}^0_1 \tilde{\chi}^+_1 \rightarrow n_1 \bar{e}_1$ & 2.4\% & 2\% & 1\% & $<$1\% & 2\% \\
$\tilde{\chi}^0_1 \tilde{\chi}^+_1 \rightarrow n_2 \bar{e}_2$ & 2.4\% & 2\% & 1\% & $<$1\% & 2\% \\
$\tilde{\chi}^0_1 \tilde{\chi}^+_1 \rightarrow n_3 \bar{e}_3$ & 2.4\% & 2\% & 1\% & $<$1\% & 2\% \\
$\tilde{\chi}^0_1 \tilde{\chi}^+_1 \rightarrow Z W^+$ & 1.0\% & $<$1\% &  $<$1\% & $<$1\% & $<$1\% \\
$\tilde{\chi}^0_1 \tilde{\chi}^+_1 \rightarrow A W^+$ & 1.2\% & $<$1\% &  $<$1\% & $<$1\% & $<$1\% \\
$\tilde{\chi}^0_1 \tilde{\chi}^+_1 \rightarrow h_1 W^+$ & 0.6\% & $<$1\% &  $<$1\% & $<$1\% & $<$1\% \\
$\tilde{\chi}^0_2 \tilde{\chi}^+_1 \rightarrow t\bar{b}$ & 6.1\% & 4\% & 5\% & $<$1\% & 5\% \\
$\tilde{\chi}^0_2 \tilde{\chi}^+_1 \rightarrow u\bar{d}$ & 5.9\% & 3\% & 5\% & $<$1\% & 5\% \\
$\tilde{\chi}^0_2 \tilde{\chi}^+_1 \rightarrow c\bar{s}$ & 6.9\% & 3\% & 5\% & $<$1\% & 5\% \\
$\tilde{\chi}^0_3 \tilde{\chi}^+_1 \rightarrow t\bar{b}$ & $<$1\% & $<$1\% & 4\% & $<$1\% & 3\% \\
$\tilde{\chi}^0_3 \tilde{\chi}^+_1 \rightarrow u\bar{d}$ & $<$1\%  & $<$1\% & 4\% & $<$1\% & 3\% \\
$\tilde{\chi}^0_3 \tilde{\chi}^+_1 \rightarrow c\bar{s}$ & $<$1\% & $<$1\% & 4\% & $<$1\% & 3\% \\
$\tilde{\chi}^0_2 \tilde{\chi}^+_1 \rightarrow n_1 \bar{e}_1$ & 2.1\% & 1\% & 2\% & $<$1\% & 2\% \\
$\tilde{\chi}^0_2 \tilde{\chi}^+_1 \rightarrow n_2 \bar{e}_2$ & 2.1\% & 1\% & 2\% & $<$1\% & 2\% \\
$\tilde{\chi}^0_2 \tilde{\chi}^+_1 \rightarrow n_3 \bar{e}_3$ & 2.1\% & 1\% & 2\% & $<$1\% & 2\% \\
$\tilde{\chi}^0_3 \tilde{\chi}^+_1 \rightarrow n_1 \bar{e}_1$ & $<$1\% & $<$1\% & 1\% & $<$1\% & 1\% \\
$\tilde{\chi}^0_3 \tilde{\chi}^+_1 \rightarrow n_2 \bar{e}_2$ & $<$1\% & $<$1\% & 1\% & $<$1\% & 1\% \\
$\tilde{\chi}^0_3 \tilde{\chi}^+_1 \rightarrow n_3 \bar{e}_3$ & $<$1\% & $<$1\% & 1\% & $<$1\% & 1\% \\
$\tilde{\chi}^0_1 \tilde{\chi}^0_1 \rightarrow W^+ W^-$ & 1.8\% & 2\% & 1\% & $<$1\% & 2\% \\
$\tilde{\chi}^0_1 \tilde{\chi}^0_3 \rightarrow W^+ W^-$ & $<$1\% & $<$1\% & 1\% & $<$1\% & $<$1\% \\
$\tilde{\chi}^0_1 \tilde{\chi}^0_1 \rightarrow Z Z$ & 1.5\% & 1\% &  $<$1\% & $<$1\% & 2\% \\
$\tilde{\chi}^0_1 \tilde{\chi}^0_1 \rightarrow t\bar{t}$ & 0.0\% & $<$1\% &  $<$1\% & 12\% & 1\% \\
$\tilde{\chi}^0_1 \tilde{\chi}^0_1 \rightarrow b\bar{b}$ & $<$1\% & 22\%  &  $<$1\% & 80\% & $<$1\% \\
$\tilde{\chi}^0_1 \tilde{\chi}^0_1 \rightarrow e_3\bar{e}_3$ & $<$1\% & 1\%  &  $<$1\% & 5\% & $<$1\% \\
$\tilde{\chi}^0_1 \tilde{\chi}^0_2 \rightarrow d\bar{d}$ & 2.2\% & 1\% & 1\% & $<$1\% & 1\% \\
$\tilde{\chi}^0_1 \tilde{\chi}^0_2 \rightarrow s\bar{s}$ & 2.2\% & 1\% & 1\% & $<$1\% & 1\% \\
$\tilde{\chi}^0_1 \tilde{\chi}^0_2 \rightarrow b\bar{b}$ & 2.2\% & 2\% & 1\% & $<$1\% & 1\% \\
$\tilde{\chi}^0_1 \tilde{\chi}^0_2 \rightarrow t\bar{t}$ &$<$1\%  &$<$1\%  & $<$1\% & $<$1\% & 1\% \\
$\tilde{\chi}^0_3\tilde{\chi}^0_2 \rightarrow d\bar{d}$ & $<$1\% & $<$1\% & 1\% & $<$1\% & $<$1\% \\
$\tilde{\chi}^0_3\tilde{\chi}^0_2 \rightarrow s\bar{s}$ & $<$1\% & $<$1\% & 1\% & $<$1\% & $<$1\% \\
$\tilde{\chi}^0_3\tilde{\chi}^0_2 \rightarrow b\bar{b}$ & $<$1\% & $<$1\% & 1\% & $<$1\% & $<$1\% \\
$\tilde{\chi}^0_1 \tilde{\chi}^0_2 \rightarrow u\bar{u}$ & 1.7\% & 1\% &  $<$1\% & $<$1\% & 1\% \\
$\tilde{\chi}^0_1 \tilde{\chi}^0_2 \rightarrow c\bar{c}$ & 1.7\% & 1\% &  $<$1\% & $<$1\% & 1\% \\
$\tilde{\chi}^0_3 \tilde{\chi}^0_2 \rightarrow u\bar{u}$ & 1.7\% & 1\% &  $<$1\% & $<$1\% & $<$1\% \\
$\tilde{\chi}^0_3 \tilde{\chi}^0_2 \rightarrow c\bar{c}$ & 1.7\% & 1\% &  $<$1\% & $<$1\% & $<$1\% \\
$\tilde{\chi}^0_2 \tilde{\chi}^0_2 \rightarrow W^+ W^-$ & 1.1\% & $<$1\% & $<$1\%  & $<$1\% & $<$1\% \\
$\tilde{\chi}^+_1 \tilde{\chi}^-_1 \rightarrow W^+ W^-$ & 2.7\% & 1\% & 2\% & $<$1\% & 2\% \\
$\tilde{\chi}^+_1 \tilde{\chi}^-_1 \rightarrow u\bar{u}$ & 2.1\% & 1\% & 2\% & $<$1\% & 2\% \\
$\tilde{\chi}^+_1 \tilde{\chi}^-_1 \rightarrow c\bar{c}$ & 2.1\% & 1\% & 2\% & $<$1\% & 2\% \\
$\tilde{\chi}^+_1 \tilde{\chi}^-_1 \rightarrow t\bar{t}$ & 2.1\% & 1\% & 2\% & $<$1\% & 2\% \\
$\tilde{\chi}^+_1 \tilde{\chi}^-_1 \rightarrow d\bar{d}$ & 1.4\% & $<$1\% & 1\% & $<$1\% & 1\% \\
$\tilde{\chi}^+_1 \tilde{\chi}^-_1 \rightarrow s\bar{s}$ & 1.4\% & $<$1\% & 1\% & $<$1\% & 1\% \\
$\tilde{\chi}^+_1 \tilde{\chi}^-_1 \rightarrow b\bar{b}$ & 1.3\% & 4\% & 1\% & $<$1\% & 1\% \\
$\tilde{\chi}^+_1 \tilde{\chi}^-_1 \rightarrow e_1 \bar{e}_1$ & 1.1\% & $<$1\% &  $<$1\% & $<$1\% & $<$1\% \\
$\tilde{\chi}^+_1 \tilde{\chi}^-_1 \rightarrow e_2 \bar{e}_2$ & 1.1\% & $<$1\% &  $<$1\% & $<$1\% & $<$1\% \\
$\tilde{\chi}^+_1 \tilde{\chi}^-_1 \rightarrow e_3 \bar{e}_3$ & 1.1\% & $<$1\% &  $<$1\% & $<$1\% & $<$1\% \\
\hline
\end{tabular}
\caption{The co-annihilation channels that contribute to $(\Omega h^2)^{-1}$ for the five benchmarks points chosen in this study. There are many other contributing channels taking the total up to 100\% for each benchmark, but for the sake of brevity they are not included in this table if they do not contribute at least 1\% for at least one benchmark point.}
\label{coannihilations1}
\end{center}
\end{table}

\section*{Acknowledgements}
We gratefully acknowledge the help of Jonathan Hall in providing the
modified {\tt micrOMEGAs} program with the E6SSM model file and
spectrum generator embedded.  PA would like also like to thank Steve
King for early discussions about the viability of this project, Roman
Nevzorov for very illuminating discussions about possible dark matter
candidates in $E_6$-inspired models and Andrew Fowlie for reading
through a draft of the manuscript and providing useful feedback. This
work was supported by the University of Adelaide, Monash University
and the Australian Research Council through the ARC Centre of
Excellence for Particle Physics at the Terascale (CoEPP)
(CE110001104). MJW is supported by the Australian Research Council
Future Fellowship FT14010024

\end{document}